\documentclass[twocolumn,english,prb]{revtex4}
\usepackage[T1]{fontenc}
\usepackage[latin9]{inputenc}
\usepackage[letterpaper]{geometry}
\usepackage{amssymb}
\usepackage{graphicx}
\usepackage{esint}

\makeatletter

\@ifundefined{textcolor}{}
{%
 \definecolor{BLACK}{gray}{0}
 \definecolor{WHITE}{gray}{1}
 \definecolor{RED}{rgb}{1,0,0}
 \definecolor{GREEN}{rgb}{0,1,0}
 \definecolor{BLUE}{rgb}{0,0,1}
 \definecolor{CYAN}{cmyk}{1,0,0,0}
 \definecolor{MAGENTA}{cmyk}{0,1,0,0}
 \definecolor{YELLOW}{cmyk}{0,0,1,0}
}

\makeatother

\usepackage{babel}

\begin{document}

\title{Pathways to faceting of vesicles}
\author{Mark J.~Bowick}
\email{bowick@phy.syr.edu}
\affiliation{Department of Physics and Soft Matter Program, Syracuse University, Syracuse, New York 13244,
USA}
\author{Rastko Sknepnek}
\email{sknepnek@gmail.com}
\affiliation{Department of Physics and Soft Matter Program, Syracuse University, Syracuse, New York 13244,
USA}

%Please do not change this text.
\begin{abstract}
The interplay between geometry, topology and order can lead to geometric
frustration that profoundly affects the shape and structure of a curved
surface. In this commentary we show how frustration in this context can result in the faceting
of elastic vesicles. We show that, under the right conditions, an assortment
of regular and irregular polyhedral structures may be the low energy
states of elastic membranes with spherical topology. In particular, we show how topological
defects, necessarily present in any crystalline lattice confined to spherical topology, naturally lead 
to the formation of icosahedra in a homogeneous elastic vesicle. Furthermore, we show that introducing
heterogeneities in the elastic properties, or allowing for non-linear bending response of a homogeneous system,
 opens non-trivial pathways to the formation of faceted, yet non-icosahedral, structures.
\end{abstract}
\maketitle

\section{Introduction}

Although most of the lipid bilayer vesicles found in biological systems are smooth one often observes faceted
 structures as well. Prominent examples in nature are the protein capsids of many large viruses, 
the protein-enclosed bacterial organelles known as carboxysomes\cite{Iancu2010}
and the square-shaped bacteria found in saturated brine pools.\cite{walsby1980square}
On the experimental side faceted forms are found in small phosphatidylcholine
vesicles below the freezing temperature,\cite{blaurock1979small}
some cationic vesicles,\cite{marques1999interactions,antunes2007mechanisms}
cross-polymerized vesicles,\cite{Greenfield2009} vesicles assembled from a mixture of anionic and
 cationic amphiphiles,\cite{Dubois2001,antunes2007mechanisms,leung2012molecular}
vesicles formed from block copolymers with liquid-crystalline side
chains,\cite{jia2009smectic,xu2009self,jia2011smectic} giant fullerenes,\cite{lamb1992extraction}
and gold nanocages.\cite{Hu2007} While diverse in their structure, all these systems share the 
common feature of possessing pathways to lower the total energy by forming stable facets. Over the past
four decades substantial research effort has been devoted to understanding
the shapes of smooth vesicles. Only recently has there been a growing interest in developing a
 theoretical understanding of faceted structures with their sharp edges and singular vertices. 

To understand the origin of faceting it is essential to understand the physics of smooth vesicles. 
Early in evolution it became clear that to construct complex life it was necessary to be able
to hierarchically compartmentalize and separate different parts of an organism. 
The compartmentalization was achieved with the aid of vesicles made of molecularly-thin membranes. 
Biological membranes are typically phospholipid bilayer structures embedded with a variety of membrane
and transmembrane proteins.\cite{Alberts2002} The membrane thickness is typically $\sim5\mathrm{nm}$  
while the lateral dimensions range from hundreds of nanometers to microns. Given this large aspect
 ratio such membranes can be treated as quasi-two-dimensional. Vesicular membranes are not limited, however, to 
biological systems. Nanometer-sized and micron-sized containers can be constructed in the laboratory using 
lipid molecules,\cite{bangham1964negative,Veatch2003} block-copolymers\cite{nardin2000polymerized} and even,
 as in the case of gold nanocages, inorganic compounds. 

While the detailed structure and function of biological membranes is exceedingly complex and still
 to be fully understood, much insight into their physical properties can be gained by constructing simple mesoscopic
models. Helfrich,\cite{Helfrich1973} in a seminal paper, argued that the quasi-two-dimensional nature
 of membranes implies that their dominant deformation is bending, corresponding to changing shape in space.    
He constructed a free energy functional that relates the bending energy to the integral
of the square of the mean curvature, $H$, and to the Gaussian curvature,
$K$, i.e. 
\begin{equation}
E_{bend}=\int dS\left[2\kappa\left(H-H_{0}\right)^{2}+\kappa_{G}K\right],\label{eq:Helfriech}
\end{equation}
where $H_{0}$ is the spontaneous curvature and $\kappa$ and $\kappa_{G}$
are respectively the bending and Gaussian rigidities which themselves depend on the molecular details. 
The Helfrich free energy describes fluid membranes with free lateral diffusion as there is no energy cost
 for shear deformations. Fluidity is important for biological membranes as it 
ensures that small apertures formed by transmembrane transport, or from damage of some kind, can be closed 
quickly. It also allows structural reorganizations within the membrane itself, as seen, for example,
 in lipid rafts.\cite{munro2003lipid}

Homogeneous fluid vesicles are typically smooth since molecules quickly rearrange themselves to remove any
 stress that would be generated by the formation of sharp corners or edges. Sharp structures correspond
to regions with very high or diverging curvature\cite{Witten2007} and are therefore energetically costly. 
While in real systems the divergence is cut off at some molecular length scale, the energy penalty for
forming sharp bends is comparable to molecular energies and far exceeds the energy cost 
($\sim k_{B}T$) of long-wavelength undulations. To form sharp structures, therefore,
there must be a mechanism that either reduces the energy cost associated with the formation of bends 
or that inhibits the flow of molecules or imposes additional ``in-vesicle'' order (e.g. liquid
crystalline) that would select a non-smooth underlying geometry. The first mechanism can be achieved by 
introducing asymmetry in the composition of the two leaflets of the bilayer induced by a phase segregation
 of the two molecular species, thus resulting in local non-zero spontaneous curvature\cite{gonzalez2007isolated}
or by segregating excess amphiphiles along long ridges and inducing
a spontaneous curvature that is commensurate with the dihedral angle
between two faces.\cite{Dubois2001,Haselwandter2010} In this paper
we discuss mechanisms that lead to faceting of elastic
vesicles. An interesting account of faceting of giant vesicles in
the rippled gel phase $P_{\beta'}$ by osmotic deflation has recently
been given by Quemeneur, \emph{et al}.\cite{quemeneur2012gel}

This paper is organized as follows. In Section \ref{sub:icosahedra}
we discus pathways to the faceting of elastic membranes
that describe viral capsids or bilayer systems cooled below the gelation
transition or for which the molecules are cross-polymerized (i.e.
tethered to each other). In particular, we show, in Section \ref{sub:icosahedra},
how topological defects can act as the seeds of buckling to
icosahedra, thus providing a simple explanation for the observation that
viruses with icosahedral symmetry are spherical for small sizes but well-faceted icosahedra 
for larger sizes. In Section \ref{sub:multi-component} we show how the presence of
elastic inhomogeneities also can lead to faceting to regular and
irregular polyhedra quite different from icosahedra. Finally,
in Section \ref{sub:critical-curvature}, we argue that faceting can
also occur in homogeneous elastic vesicles if one allows for a non-linear
bending response in terms of a critical curvature.

\section{Faceting of elastic membranes}

\label{sec:elastic_membranes}

\subsection{Buckling into icosahedra}

\label{sub:icosahedra}One of the hallmarks of viruses is the regular
structure of their capsids. The unit building blocks of capsids are
capsomeres, robust protein complexes, $\approx10\mathrm{nm}$ in diameter,
that tile the entire capsid in a regular (crystalline) array. The lattice structures of
icosahedral viruses can be described in terms of pairs of non-negative
integers, $\left(p,q\right)$ that form a $T-$number, $T=p^{2}+q^{2}+pq.$\cite{Caspar62}
The existence of a lattice combined with the spherical topology is
at the center of the argument for size-driven buckling into icosahedra.
To properly explain the mechanism behind this buckling phenomenon
we need first to briefly review the complex question of the existence
of long-range order on curved surfaces.

Constructing a regular lattice in the plane (or any other flat surface) is straightforward. 
The simplest example is the ground state of identical point-like particles in an external 
confining potential interacting via long range repulsive interactions, e.g. identical
charges confined to a planar region. The ground state is a triangular
lattice with each particle having exactly six equidistant nearest
neighbors and a lattice spacing that is determined by the density.
Such a state is not frustrated and is stress free. This is no longer
the case when one attempts a similar construction on the 2-sphere (the surface of a ball in $\mathbb{R}^3$).
Any attempt to map a regular lattice on to the surface of
a 2-sphere will result in excess particles. In order to build a smooth covering of the 2-sphere
it is necessary, in other words, to remove parts of the original planar lattice. As
a result, even in the ground state any crystalline lattice on a 2-sphere
will necessarily have a finite number of sites that have coordination 
different from six, i.e. so called topological defects. Starting
from Euler's polyhedral formula, which states that for any polyhedron
$V-E+F=\chi$, where $V$ is the number of vertices, $E$ is the number
of edges, $F$ is the number of faces, and $\chi$ is the Euler characteristic
of the polyhedron, one can show that the total number of vertices
of a crystalline lattice on a sphere has to satisfy the following
relation:\cite{Sachdev1984} $\sum_{z}\left(6-z\right)N_{z}=6\chi=12$.
Here $N_{z}$ is the total number of sites with coordination $z$ and we
have explicitly used that $\chi=2$ for a 2-sphere. If we restrict ourselves to coordination numbers 
$z=5,6$ or $z=7$ (physically coordination numbers are near 6), the configuration
with the lowest number of defects on a sphere has $N_{5}=12$, i.e.
twelve five-fold coordinated sites embedded in a regular triangular
lattice. We note that, somewhat counterintuitively, this configuration
is not always the ground state of the system as shown by Bowick, \emph{et
al.}\cite{Bowick2000} For sufficiently large total number of particles, and reasonable
 defect core energies, the total energy can be lowered by forming pairs of five- and seven-fold defects
that emanate from each of the twelve five-fold coordinated sites.
In the following discussion we will ignore this complication and assume
that only twelve five-fold defects are present.

Despite their appearance five-fold disclination defects are not local
since they owe their existence to removing whole sections of a planar
lattice so that it conforms to a 2-sphere. As such they are endowed with considerable elastic energy. 
To estimate how the elastic energy of a disclination depends on its size we assume that
it is surrounded by a Hookean elastic medium. The elastic energy of
a two-dimensional crystal is\cite{LandauVol7} $E_{el}=\int d^{2}\vec{r}A^{ijkl}u_{ij}u_{kl}$,
where $u_{ij}\left(\vec{r}\right)=\frac{1}{2}\left(\partial_{i}u_{j}+\partial_{j}u_{i}+\partial_{k}u_{i}\partial_{k}u_{j}\right)$
is the strain tensor field, with $u_{i}\left(\vec{r}\right)$ ($i=1,2)$
being the components of the displacement field, and $A^{ijkl}$ is the
material-dependent elastic tensor. For a hexagonal lattice, the elastic
tensor has only two independent components,\cite{LandauVol7} i.e.
two Lam{\' e} coefficients $\mu$ and $\lambda$ and the elastic energy
simplifies to\cite{LandauVol7} 
\begin{equation}
E_{el}=\frac{1}{2}\int d^{2}\vec{r}\left(\lambda\left(u_{ii}\right)^{2}+2\mu u_{ij}u_{ij}\right),\label{eq:elastic-energy}
\end{equation}
and we assume summation over pairs of repeated indices.

A planar five-fold disclination defect can be constructed by removing
a $60^{\circ}$ wedge of a triangular lattice and sewing
the two cut edges together. A detailed calculation\cite{Seung88}
finds that the stretching energy of a five-fold disclination is $E_{el}^{5-fold}=\left(Y/32\pi\right)R^{2}$,
where $Y$ is the two-dimensional Young's modulus and $R$ is the system size.

If the disclination is allowed to buckle out of the plane it can lower
its energy by forming a conical structure with the apex at the disclination.
In this case the stretching energy is relieved at the expense of
a bending penalty. The bending penalty is given by\cite{koiter1959consistent,Seung88}
\begin{equation}
E_{bend}=\int dS\left(2\kappa H^{2}+\kappa_{G}K\right).\label{eq:bending_energy}
\end{equation}
It can be shown\cite{Seung88} that the bending energy of such a conical
structure is $E_{bend}^{\left(5\right)}\approx\left(\pi/3\right)\kappa\log\left(R/R_{b}\right)+\frac{1}{32\pi}YR_{b}^{2}$,
where $R_{b}\approx\sqrt{154\kappa/Y}$. By comparing the expressions
for the stretching and bending energies it is clear that for small radii
stretching energy is smaller than bending energy and the disclination
will remain flat at the expense of stretching its surrounding. As
$R$ increases, the logarithm grows less rapidly than $R^{2}$ and
it becomes energetically favorable to buckle into a cone. A detailed
calculation\cite{Seung88} shows that the buckling transition occurs when the dimensionless ratio, 
the so-called F{\" o}ppl-von K{\' a}rm{\' a}n
or FvK number, $\gamma=\frac{YR^{2}}{\kappa}\gtrsim154$.

\begin{figure}
\begin{minipage}[t]{0.49\columnwidth}%
\begin{center}
\includegraphics[scale=0.18]{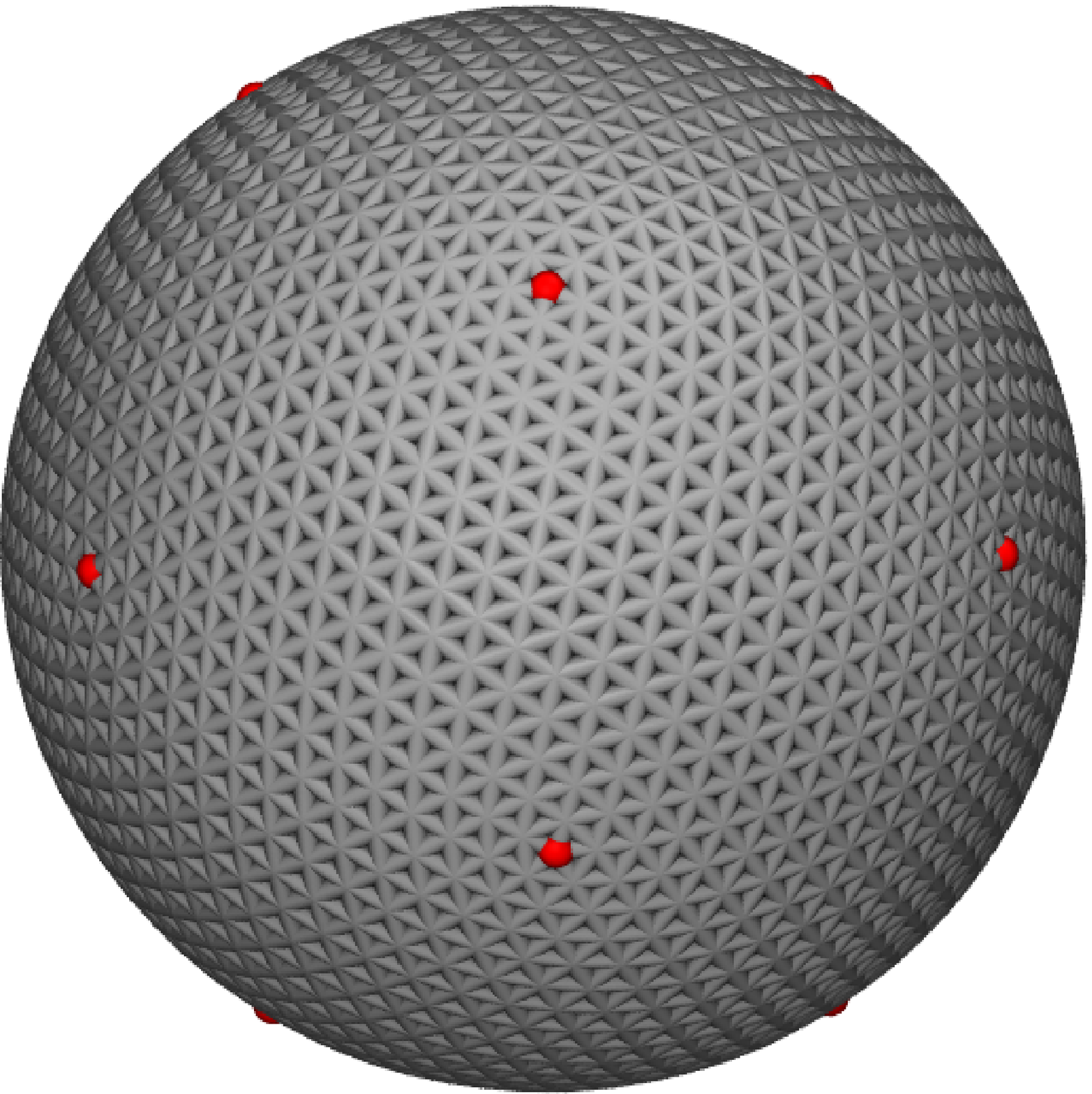} 
\par\end{center}

\begin{center}
(a) 
\par\end{center}%
\end{minipage}%
\begin{minipage}[t]{0.49\columnwidth}%
\begin{center}
\includegraphics[scale=0.18]{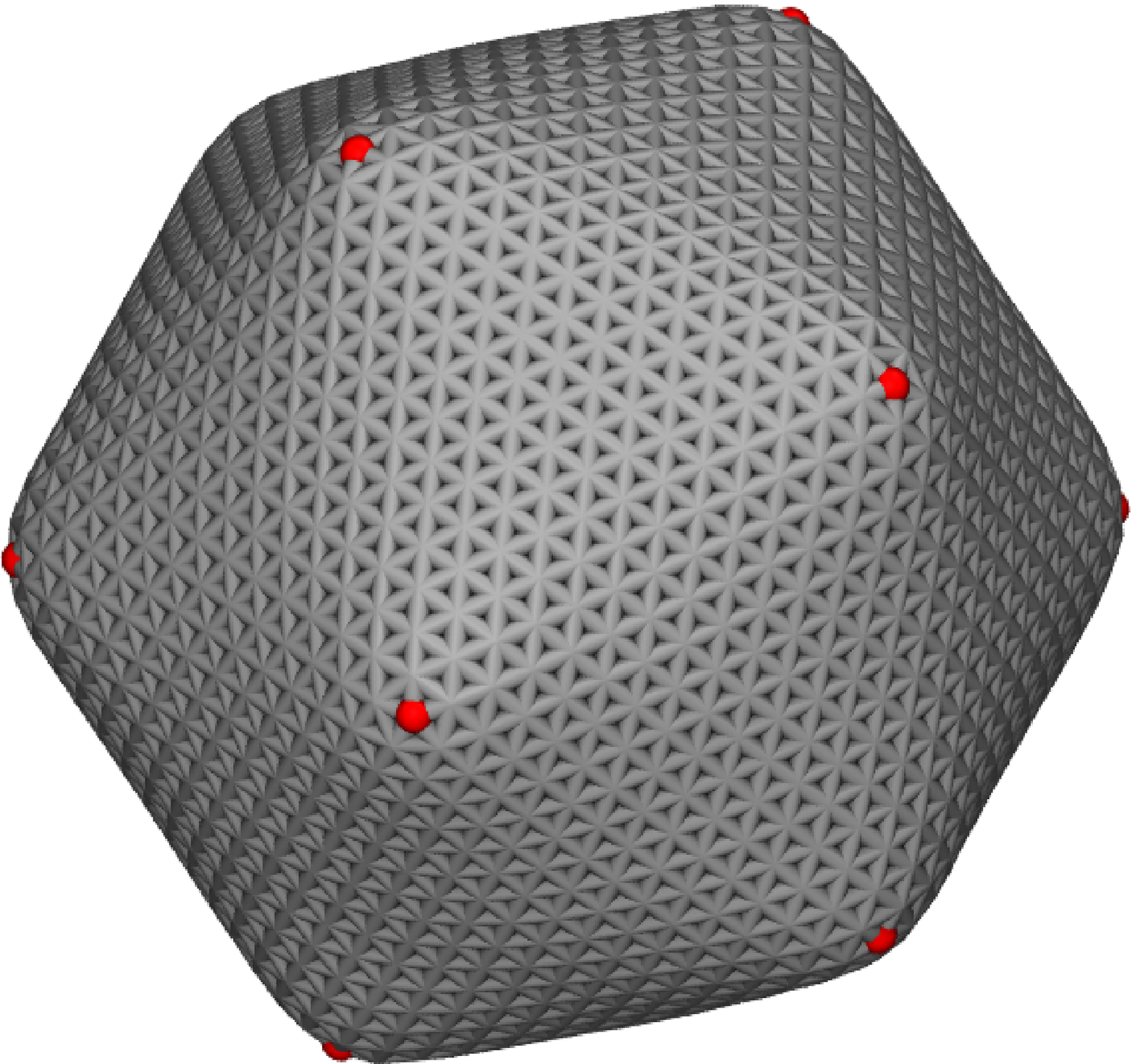} 
\par\end{center}

\begin{center}
(b) 
\par\end{center}%
\end{minipage}

\caption{The transition is triggered if the dimensionless
FvK number $\gamma$ exceeds a critical value $\gamma\sim10^{2}$.\cite{Lidmar03}
Simulated structures are constructed using the $\left(p,q\right)=\left(12,3\right)$
icosadeltahedral triangulation with $\approx2\times10^{3}$ vertices and
radius $R\approx11.5l_{0}$, with $l_{0}=1$ being the average lattice
spacing. (a) $\gamma\approx170$ and (b) $\gamma\approx3500$. Red
dots designate five-fold defects. All snapshots are generated with
the Visual Molecular Dynamics (VMD) package\cite{Humphrey96} and
rendered with the Tachyon ray-tracer.\cite{Stone98} Disclination-driven buckling 
into an icosahedron is discussed in Ref.~\citenum{Lidmar03}. 
Results of that study are reproduced here for completeness.
 \label{fig:Disclination-driven-buckling}}
\end{figure}

Another consequence of their non-local character is that disclination
defects, much like charges, exert long-range forces on each other.
This long-range force is transmitted via the elastic deformation of the
medium with the speed of sound playing the role of the speed of light
in electrostatics. Disclinations with the same topological charge
($s=1$ for a five-fold defect) repel each other. On a 2-sphere, therefore,  
disclinations adopt a configuration that maximizes their separation;
this is the icosahedron for twelve defects.\cite{Bowick2000}

Moving from planar to spherical geometry does not change the functional
forms of the stretching and bending energies but it does renormalize
the numerical prefactors due to partial screening of the strain induced
by the non-zero Gaussian curvature of the sphere.\cite{Bowick2000}
As in the planar case, therefore, the onset of buckling depends solely
on the relative strengths of the stretching and bending energies. Since
the stretching energy grows rapidly with the radius of the sphere,
all defects buckle simultaneously into cones once a critical radius is reached and an  
icosahedral vesicle is formed.\cite{Lidmar03} In Fig.~\ref{fig:Disclination-driven-buckling} we show 
the shape of the shell below and above the buckling transition. The transition
is not sharp but rather a rounded smooth crossover, with vesicles having
FvK number $\gamma\lesssim150$ being spherical, for $200\lesssim\gamma\lesssim1500$
being noticeably buckled, and for $\gamma\gtrsim2000$ being very
sharp. For extremely large FvK numbers ($\gamma\gtrsim10^{7}$) the
vesicle is non-extensible and one observes characteristic ridge-like
structures with very interesting scaling properties, e.g. $E_{ridge}/\kappa\sim\alpha^{7/3}\left(YL^{2}/\kappa\right)^{1/6}$,
where $\alpha$ is the angle of the ridge and $L$ is its length.\cite{Lobkovsky1996,lobkovsky1997properties}

Finally we note that imposing internal pressure,\cite{Siber06} or
a volume constraint,\cite{Siber06,funkhouser2012topological} shifts the buckling transition 
to higher values of $\gamma$ but leaves the basic physics unchanged.

\subsection{Buckling of multicomponent vesicles}

\label{sub:multi-component}An implicit assumption in the discussion
thus far has been that the vesicle is made of an elastically homogenous
material. Introducing heterogeneities by allowing for spatially varying
Young's modulus and bending rigidity can lead to alternative buckling
mechanisms. The presence of heterogeneities, for example, is believed to
be responsible for the structure of the recently synthesized assemblies
of oppositely charged amphiphiles.\cite{Greenfield2009,leung2012molecular}
The structures observed are $\sim100\mathrm{nm}$ bilayer vesicles
that are faceted but not icosahedral. SAXS/WAXS measurements combined
with atomistic and detailed coarse-grained molecular dynamics simulations have
revealed the existence of strongly correlated crystalline domains
within the bilayer.\cite{leung2012molecular} A typical domain size
was found to be $\approx25\mathrm{nm}$. Simulations show that the
crystallographic axes of two neighboring domains are not mutually aligned
and that the domains are separated by narrow regions of disordered,
or even liquid, amphiphiles. It is reasonable to expect that the bilayer
will be softer along those disordered boundary regions. These regions
represent grain boundary defects with very different physical properties
from the topological defects discussed in the previous section.

To model a vesicle with inhomogeneous elastic parameters
we generalize the expressions for stretching (Eq.~(\ref{eq:elastic-energy}))
and bending (Eq.~(\ref{eq:bending_energy})) energies and allow the
two Lam{\' e} coefficients and the bending rigidity to depend on the
spatial position, i.e. $\lambda=\lambda\left(\vec{r}\right)$, $\mu=\mu\left(\vec{r}\right)$,
and $\kappa=\kappa\left(\vec{r}\right)$. We assume for simplicity
that each of these coefficients can take only two values, denoted
as ``hard'' and ``soft'' and loosely corresponding to strongly
correlated facets and grain boundary regions, respectively.

The elasticity theory of thin objects is non-linear\cite{Audoly2010}
and if one allows for spatially varying elastic parameters (even if
they have a simple binary distribution) it quickly becomes analytically
intractable. We therefore turn to numerical simulations.
The vesicle is represented as a discrete triangulated surface. In
the case of modeling viral capsids, the vertices of the triangulation
correspond to individual capsomeres. If the constitutive units are
smaller, as in the case of amphiphilic vesicles, the triangulation
can be thought of as a coarse-graining scheme that assigns discrete
elements to patches of actual molecules. Each patch is large enough
to contain a sufficient number of microscopic degrees of freedom that
molecular details are of no importance, but small enough to be considered
homogeneous.

On a triangular lattice the simplest form of the discrete stretching
energy for a multicomponent system is obtained if we assume that
the neighboring sites are connected to each other with harmonic springs
with spatially-dependent spring constants $\varepsilon_{ij}$ and
a uniform rest length $l_{0}$\cite{Seung88} 
\begin{equation}
\tilde{E}_{el}=\frac{1}{2}\sum_{\left\langle i,j\right\rangle }\varepsilon_{ij}\left(l_{ij}-l_{0}\right)^{2},\label{eq:discrete_stretch}
\end{equation}
where $l_{ij}=\left|\vec{r}_{i}-\vec{r}_{j}\right|$ is the Euclidean
distance between two sites at $\vec{r}_{i}$ and $\vec{r}_{j}$, respectively, $\varepsilon_{ij}\in\left\{ \varepsilon_{hard},\varepsilon_{soft}\right\} $,
and the sum is carried out over all pairs of nearest neighbors. It
is convenient to set $l_{0}=1$ and use it as the unit of length.

A proper discretization of the bending energy is more complicated
with a number of different approaches having been proposed in the physics, applied mathematics
and computer graphics literature.\cite{Seung88,brakke1992surface,Gompper1996,meyer2002discrete,magid2007comparison}
Different approaches vary in the level of computational complexity
and accuracy. For the present discussion it is sufficient to adopt
the simple discretization suggested by Seung and Nelson.\cite{Seung88}
The discrete bending energy is computed as 
\begin{eqnarray}
\tilde{E}_{bend} & = & \frac{1}{2}\sum_{\left\langle I,J\right\rangle }\tilde{\kappa}_{IJ}\left(\vec{n}_{I}-\vec{n}_{J}\right)^{2}\nonumber \\
 & = & \sum_{\left\langle I,J\right\rangle }\tilde{\kappa}_{IJ}\left(1-\cos\left(\theta_{IJ}\right)\right),\label{eq:discrete_bending}
\end{eqnarray}
where $\theta_{IJ}$ is the angle between the unit-length normals $\vec{n}_{I}$
and $\vec{n}_{J}$ to the neighboring triangles $I$ and $J$, and
$\tilde{\kappa}_{IJ}$ is the position dependent discrete bending
rigidity also chosen from a binary set $\left\{ \kappa_{hard},\kappa_{soft}\right\} $.
The spring constant and the discrete bending rigidity are related
to the Young's modulus and the bending rigidity of the continuum theory
by $Y=2\varepsilon/\sqrt{3}$\cite{Seung88} (with Poisson ratio
$\nu=1/3$) and $\kappa=\sqrt{3}\tilde{\kappa}/2$.\cite{Seung88,schmidt2012}

We also need to specify which element of the discrete mesh
carries which type of information. We note that discrete version of both
stretching and bending energies are most naturally defined on edges,
i.e. stretching energy is modeled by assigning a linear elastic spring
to each edge and bending energy is proportional to the dihedral angle
between two neighboring triangles sharing an edge. One is therefore tempted to 
assign component types to edges. While possible, this
is not the most convenient approach as either a vertex or a triangle,
and not an edge, is a natural discrete representation of an infinitesimal
area element in the continuum description. Defining a discrete
form of line tension energy, as discussed below, is also exceedingly
hard for component types assigned to edges. We opt rather for defining
component types on vertices. In this case a vertex $i$ is assumed
to carry a half-spring with a spring constant $\varepsilon_{i}$ which
is connected in series to the half-spring of its neighbor $j$ with
a spring constant $\varepsilon_{j}$. The spring constant of the edge
$\left(ij\right)$ is then given by $\varepsilon_{ij}^{-1}=\varepsilon_{i}^{-1}+\varepsilon_{j}^{-1}$.
For the bending energy we choose to define the bending rigidity
of an edge as the arithmetic mean of the bending rigidities assigned
to its vertices, i.e. $\kappa_{ij}=\left(\kappa_{i}+\kappa_{j}\right)/2$.
There are two side effects of defining components on vertices. First,
the spring constant of an edge shared by two vertices of the same type
is $\varepsilon_{ij}=\varepsilon_{i}/2$; and second, while the total
number of vertices of a given type is conserved, the vertex-type swaps
during the Monte Carlo optimization do not preserve the total number
of edge spring constants and bending rigidities. Extensive
tests, however, did not reveal any qualitative difference between systems with
components defined on edges and systems with components defined on
vertices or triangles. 

\begin{figure}
\begin{minipage}[t]{0.49\columnwidth}%
\begin{center}
\includegraphics[scale=0.15]{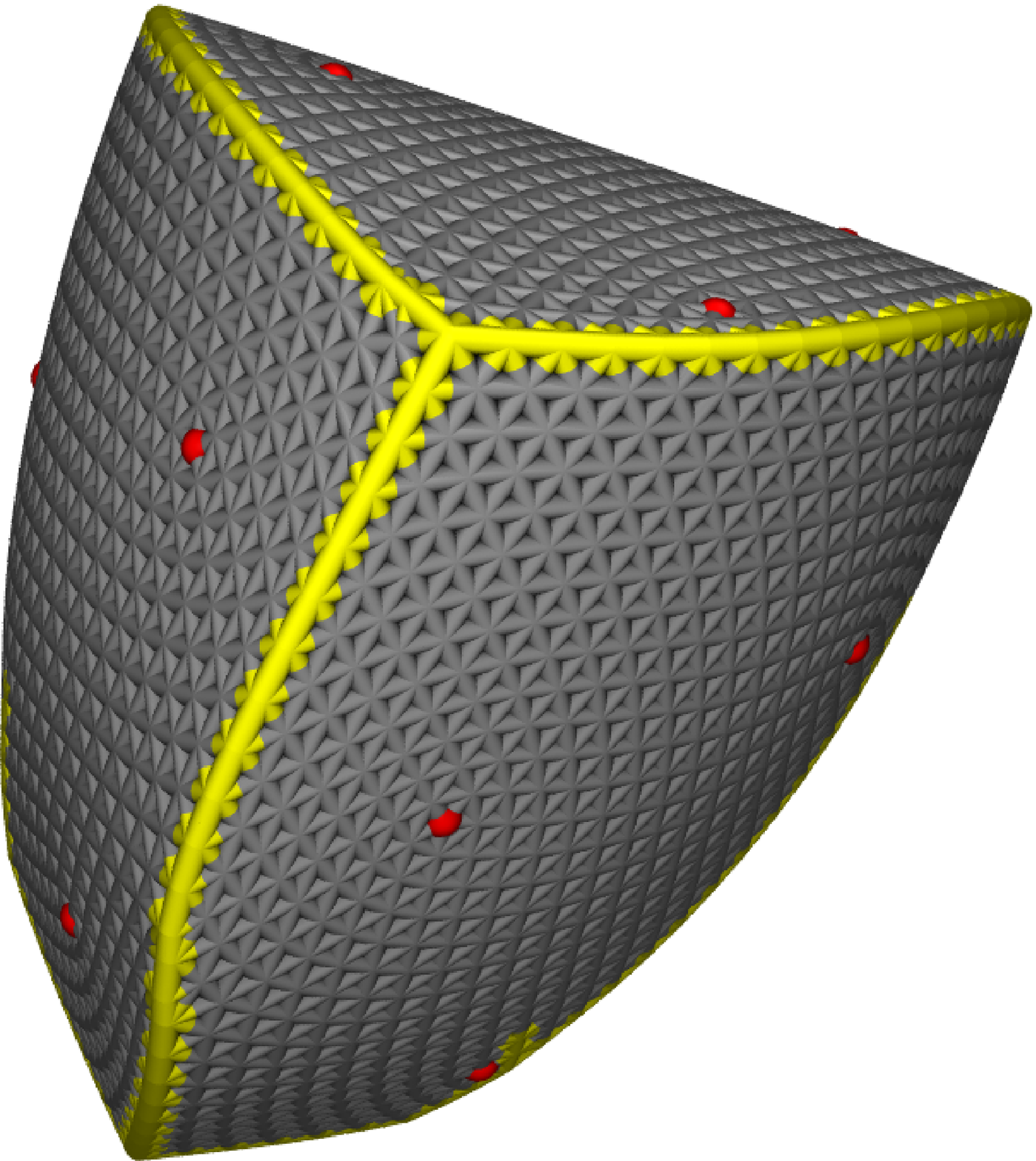} 
\par\end{center}

\begin{center}
(a) 
\par\end{center}%
\end{minipage}%
\begin{minipage}[t]{0.49\columnwidth}%
\begin{center}
\includegraphics[scale=0.15]{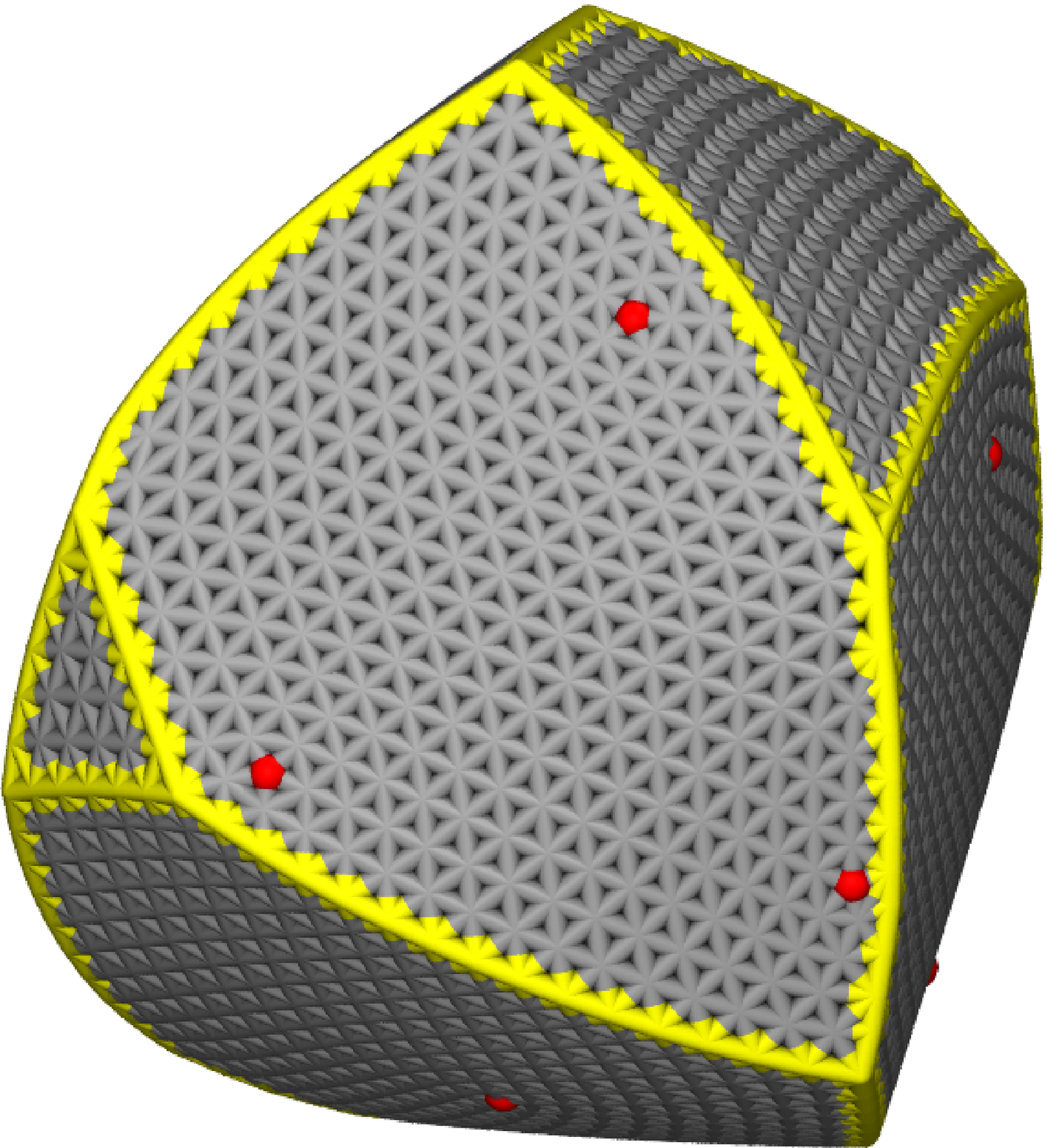} 
\par\end{center}

\begin{center}
(b) 
\par\end{center}%
\end{minipage}

\begin{minipage}[t]{0.49\columnwidth}%
\begin{center}
\includegraphics[scale=0.12]{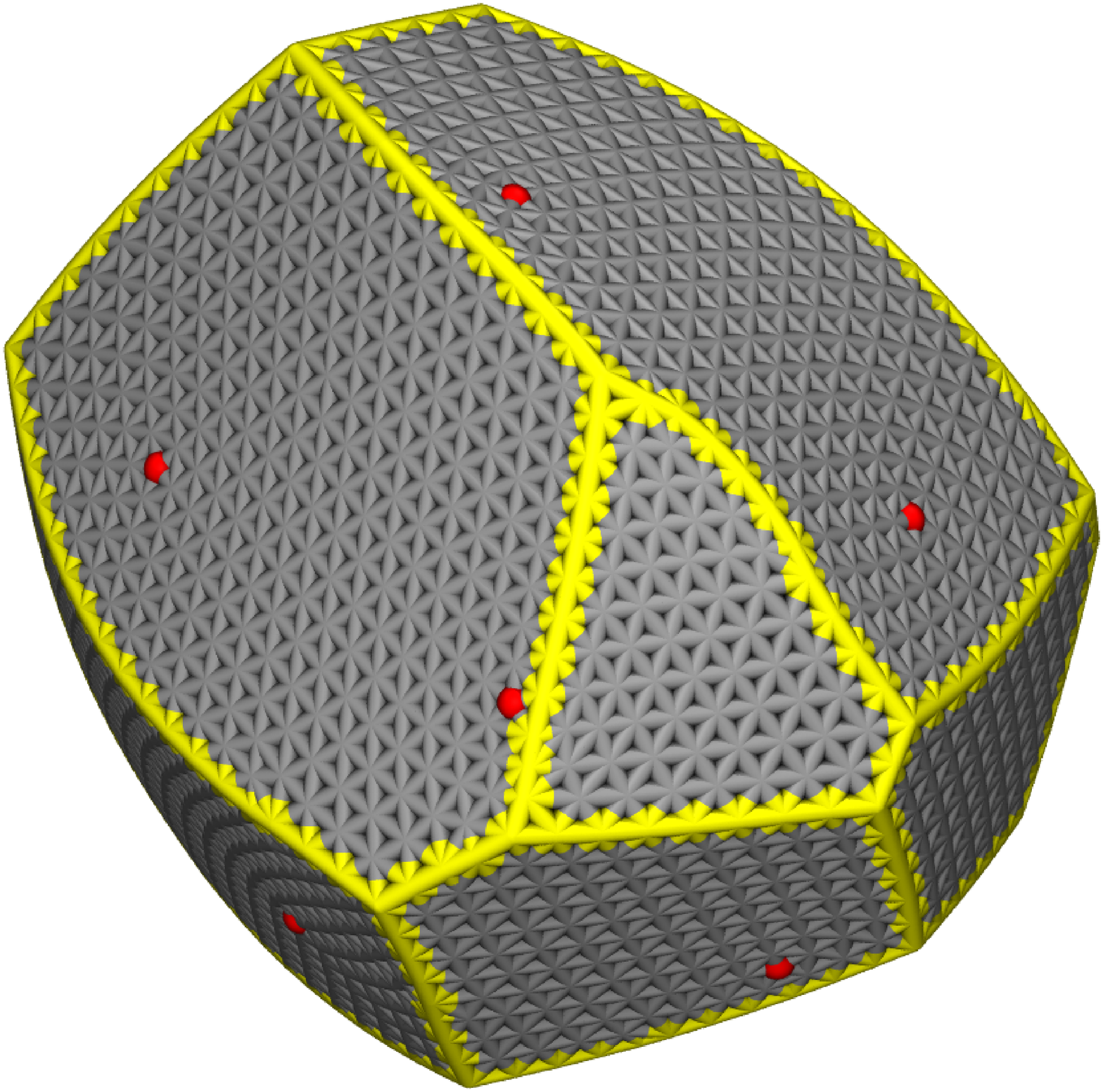} 
\par\end{center}

\begin{center}
(c) 
\par\end{center}%
\end{minipage}%
\begin{minipage}[t]{0.49\columnwidth}%
\begin{center}
\includegraphics[scale=0.14]{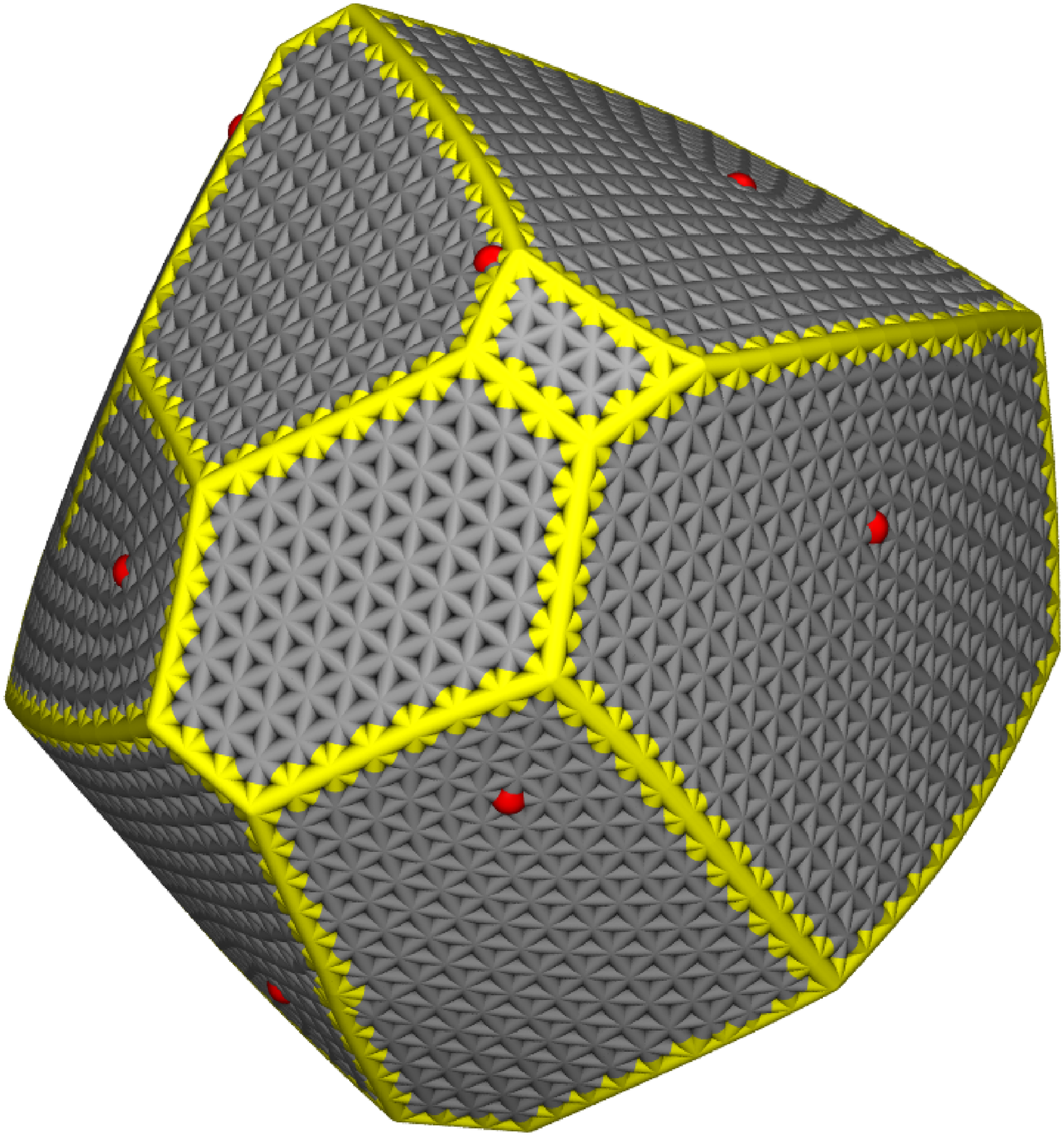} 
\par\end{center}

\begin{center}
(d) 
\par\end{center}%
\end{minipage}

\caption{Snapshots of faceted structures with 3\% (a), 4\% (b), 4.5\% (c),
and 5\% (d) of the soft component (yellow). Twelve five-fold disclination
defects are present, as required by the spherical topology, but appear
to have no influence on the position of the edges and the size of the facets.
A detailed account of faceting of two-component elastic vesicles is
presented in Ref.~\citenum{vernizzi2011platonic} and reproduced here for completeness.\label{fig:Snapshots-of-faceted}}
\end{figure}

For multicomponent vesicles we require a numerical minimization technique 
that simultaneously optimizes the position of the individual components as
well as the relative population of components. This can be achieved by using stochastic
minimization techniques, e.g. the simulated annealing Monte Carlo
method. Two components are assigned at random to vertices (or equivalently
to edges or triangles) of the initial sphere such that the total fraction
of the soft component is $f$. A Monte Carlo move has two steps: 
1) a vertex is displaced by a vector $\Delta\vec{r}$ with
components chosen at random from a uniform distribution in an interval
$\left[-\zeta l_{0},\zeta l_{0}\right]$ (typically, $\zeta\approx0.05$)
followed by 2) swap of types (\emph{soft}$\leftrightarrow$\emph{hard})
of a pair of randomly selected vertices. In both steps moves are
accepted according to Metropolis rules. During a simulation, various
cooling protocols can be applied such as linear, exponential or power-law.
We note that the annealing temperature is not the actual physical
temperature but rather a parameter that controls the acceptance rate of 
the Monte Carlo moves which increase the energy. During the component swap 
stage the component-type of a vertex (either soft or hard) is preserved so that the total 
fraction of each component is also constant. Finally we note that the component swap move is purely a convenient
simulation tool that allows sampling of the space of component arrangements {--}
it does not correspond to an actual rearrangement of the material within the vesicle which is assumed 
solid without diffusion.Throughout the simulation the triangulation is preserved {--}  no edge-flip
moves\cite{nelson2004statistical} that would create or annihilate defects are performed.

The main drawback of the simulated annealing approach is that it typically
converges very slowly once a minimum has been approached (for the studied
system sizes a minimum is typically reached after $10^{5}-10^{6}$
Monte Carlo moves) and there is no guarantee that the obtained structure
is a true global minimum, i.e. the ground state configuration. 
We thus refer to the shapes we obtain as \emph{typical}. On the other hand
the experimentally observed shapes may not be true equilibrium shapes either {--} they
are likely to depend in some part on the details of the assembly process.  

A detailed study\cite{vernizzi2011platonic} of the shapes of two-component elastic 
vesicles has found a wide variety of regular and irregular polyhedral shapes, some 
of which are shown in Fig.~\ref{fig:Snapshots-of-faceted}.
If a small amount of soft component is added to an otherwise hard
vesicle it arranges into branching lines. As the amount of soft
component is increased, these lines start to merge and form facets.
Facets are nearly flat with very sharp bends along the soft ridges.
The total number of facets increases with the amount of the soft component
until a non-universal, size-dependent concentration is reached. Beyond
this size-dependent concentration of the soft component it is no longer
favorable to generate new facets but instead the soft component is
randomly distributed inside the existing ones. Although each configuration
still contains twelve five-fold defects, it appears that the size
of the facets and the position of the soft ridges is not sensitive
to their presence (Fig.~\ref{fig:Snapshots-of-faceted}). We point
out that it is possible to entirely remove the residual strain induced
by the defects either by appropriately tuning the spring rest lengths\cite{katifori2010foldable}
or by directly working with reference and realized metrics.\cite{green2002theoretical,efrati2009elastic}
Such an approach\cite{sknepnek2012nonlinear} shows that faceting
of the soft-hard two-component vesicles occurs even if the defect
contribution to the elastic energy is removed.

\begin{figure}
\begin{minipage}[t]{0.49\columnwidth}%
\begin{center}
\includegraphics[scale=0.15]{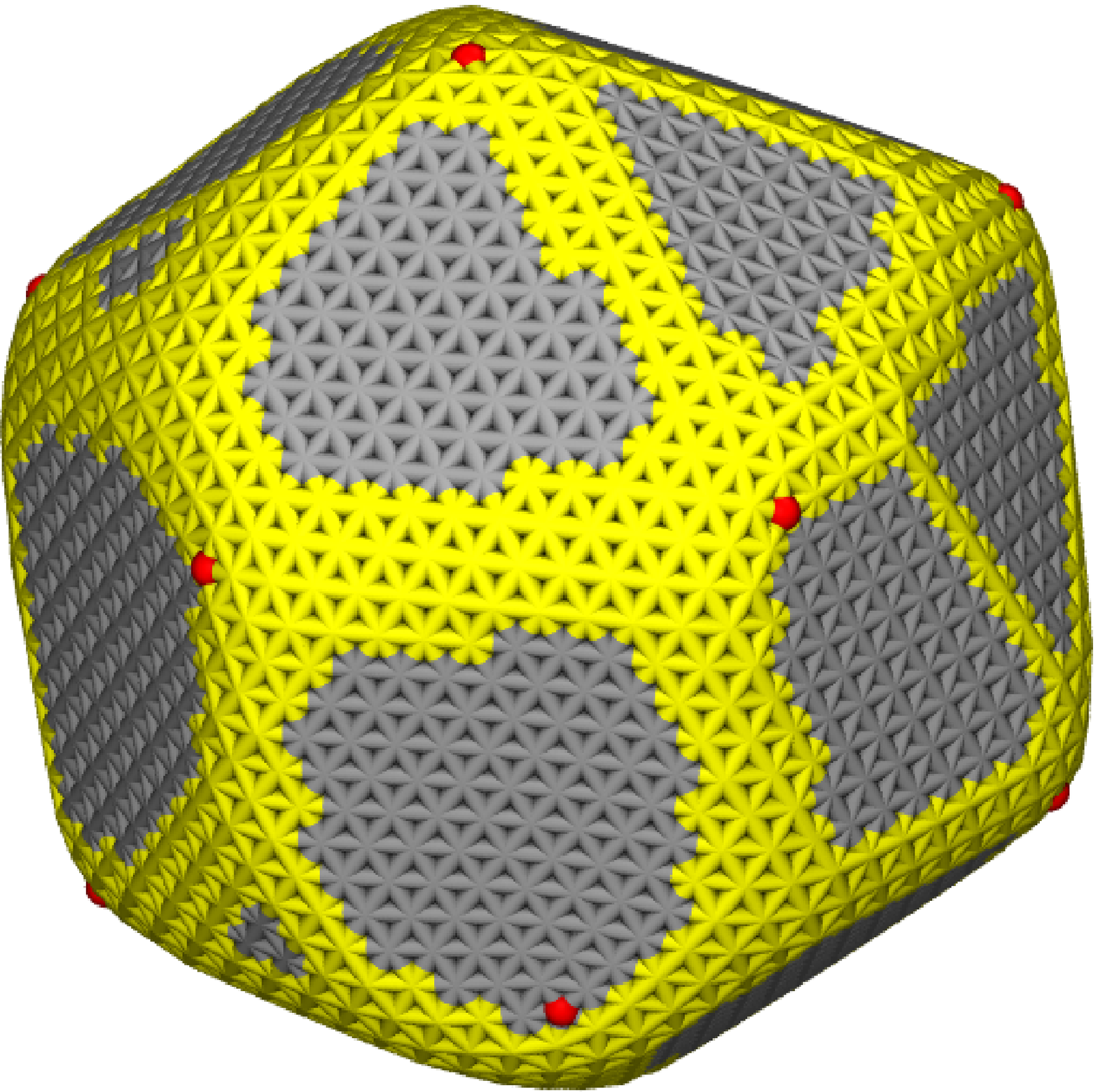} 
\par\end{center}

\begin{center}
(a) 
\par\end{center}%
\end{minipage}%
\begin{minipage}[t]{0.49\columnwidth}%
\begin{center}
\includegraphics[scale=0.15]{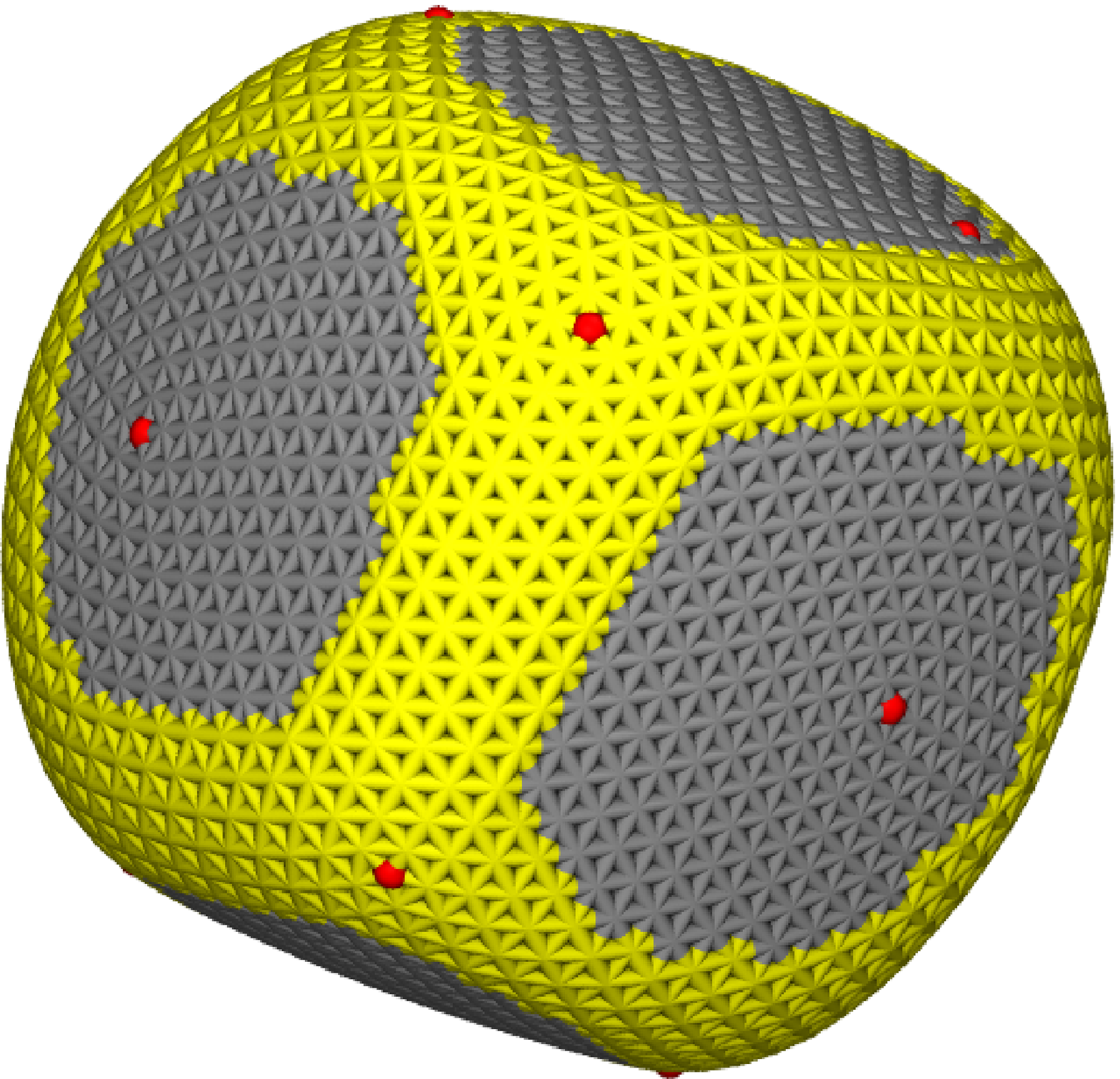} 
\par\end{center}

\begin{center}
(b) 
\par\end{center}%
\end{minipage}

\begin{minipage}[t]{0.49\columnwidth}%
\begin{center}
\includegraphics[scale=0.12]{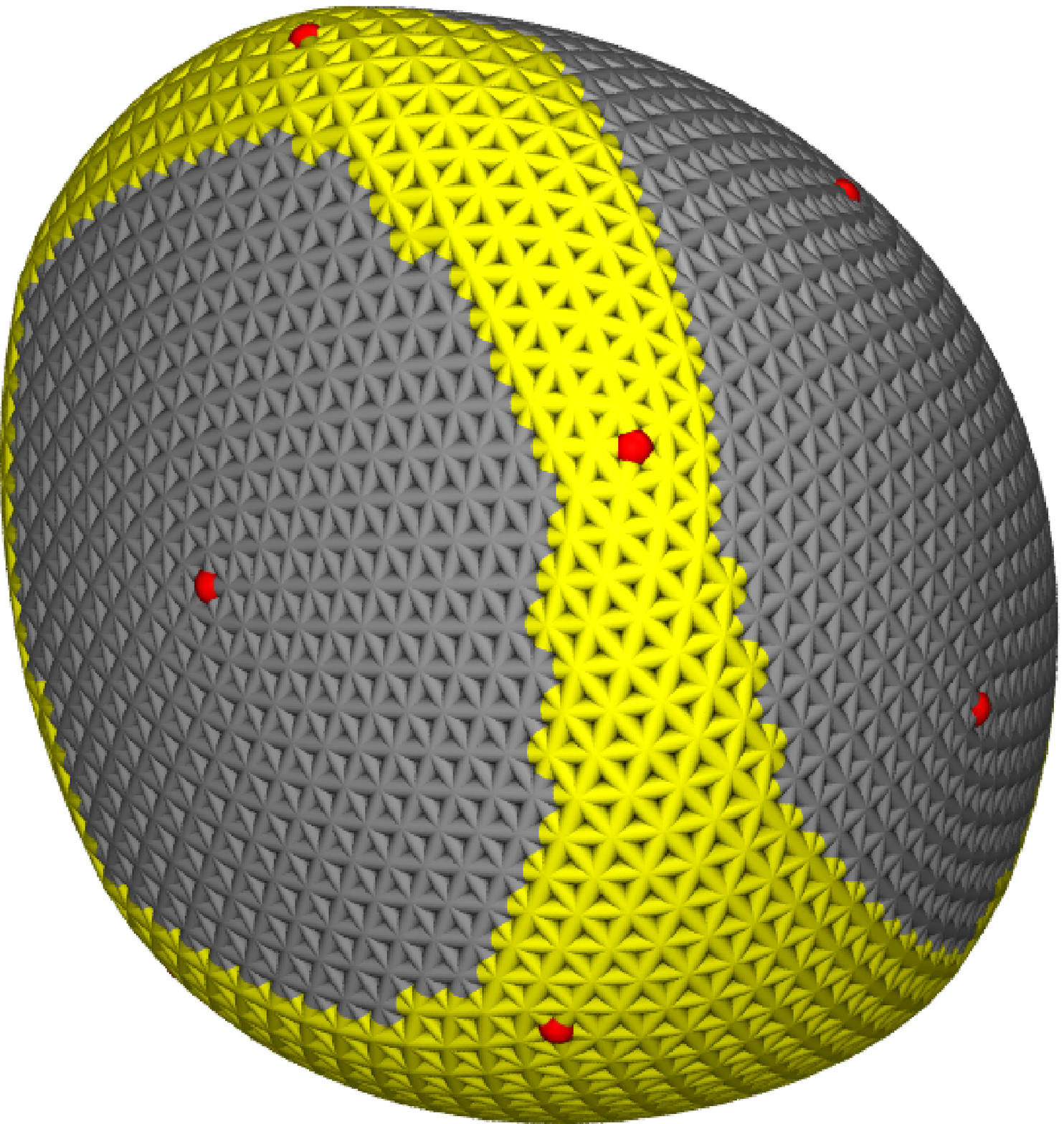} 
\par\end{center}

\begin{center}
(c) 
\par\end{center}%
\end{minipage}%
\begin{minipage}[t]{0.49\columnwidth}%
\begin{center}
\includegraphics[scale=0.14]{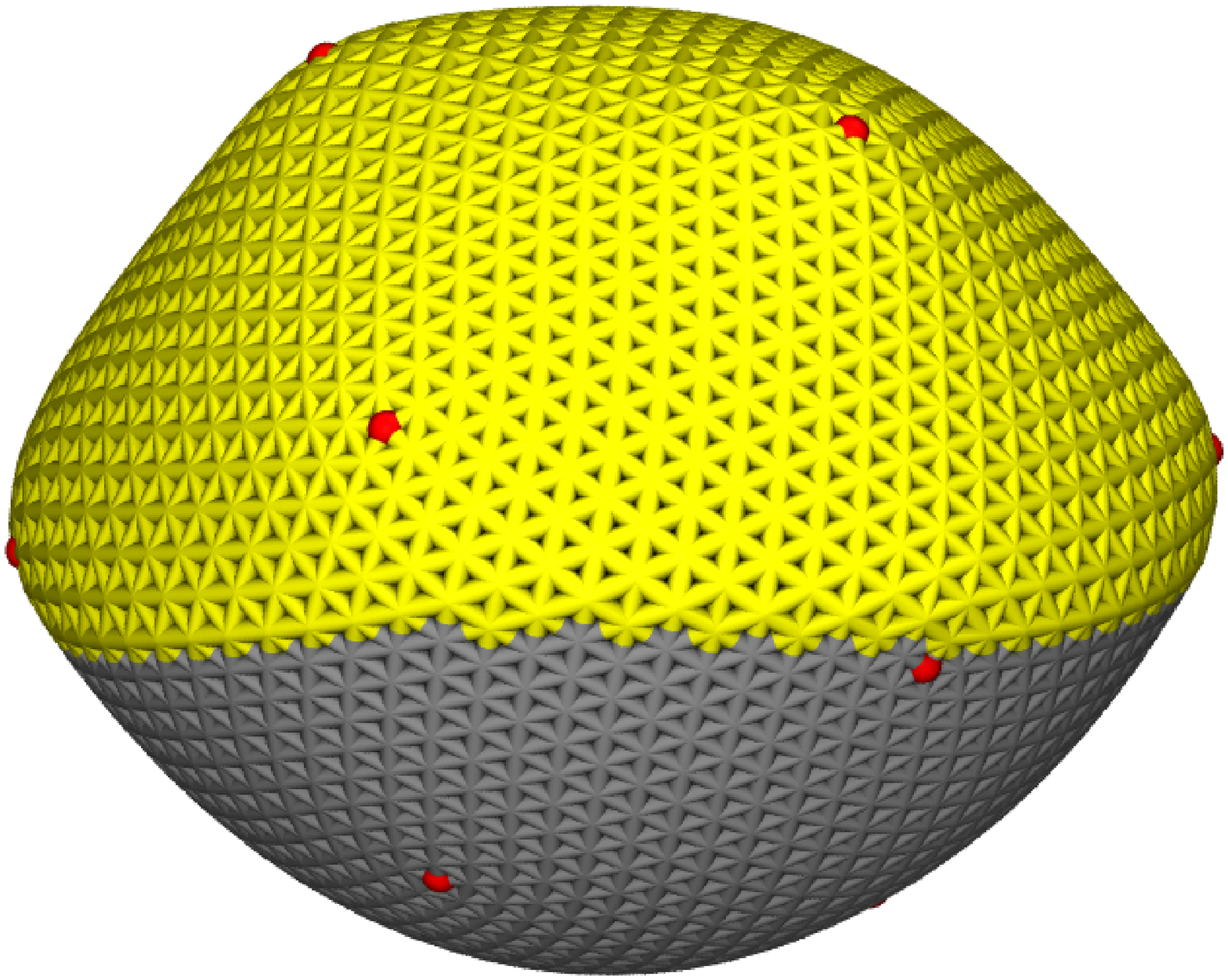} 
\par\end{center}

\begin{center}
(d) 
\par\end{center}%
\end{minipage}

\caption{Snapshots of faceted elastic vesicles with various mixing penalties
between soft (yellow) and hard (gray) components for a system with
an equal number of hard and soft vertices. The mixing penalty increases
from (a) to (d). A detailed discussion of the possible phases in two-component 
elastic vesicles with line tension is given in Ref.~\citenum{sknepnek2011buckling}.
The key results of that study are reproduced here for completeness.\label{fig:Snapshots-line-tension}}
\end{figure}

Thus far we have not assumed any mixing penalty for the two components.
The components segregated solely as a result of the bending preference.
If a mixing penalty, in the form of an effective line tension, $\Gamma$, is
introduced it competes with the preferred elastic organization of the components and 
leads to an even richer variety of shapes.\cite{sknepnek2011buckling} 
In an experimental realization of this system one can envision using the recently discovered linactant molecules,\cite{trabelsi2008linactants,trabelsi2009correlating,malone2010self}
which are two-dimensional analogues of surfactants. For weak line tension elasticity
dominates and the soft component forms highly bent ridges separating
flat hard faces. The width of these soft ridges grows as the line
tension increases. If $\Gamma$ is increased even further the faces
start to coarsen and merge together until the components fully separate
at $\Gamma l_{0}/\kappa_{hard}\approx10^{-2}$,\cite{sknepnek2011buckling}
suggesting that even a small mixing penalty can lead to
full phase segregation. A subset of shapes of two-component vesicles
with line tension is shown in Fig.~\ref{fig:Snapshots-line-tension}.

We note that line tension can also lead to very interesting shapes in two-\cite{Hu2011}
and three-component liquid vesicles.\cite{demers2012curvature}

\subsection{Faceting due to critical curvature}

\label{sub:critical-curvature} 
\begin{figure}
\begin{centering}
\includegraphics[scale=0.5]{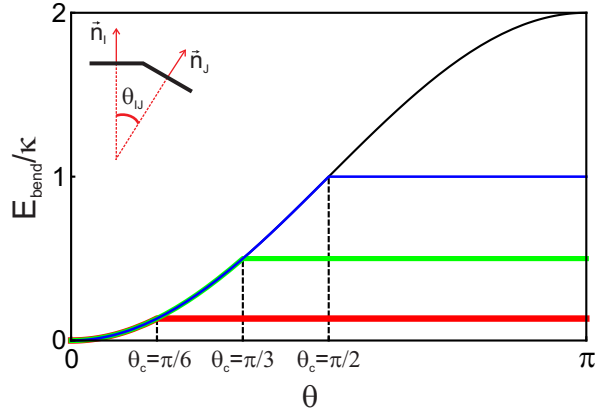} 
\par\end{centering}

\caption{Bending energy (in units of the bending rigidity $\kappa$) as a function
of the angle $\theta$ between unit-length normals to two neighboring
triangles without a critical angle (black), with $\theta_{c}=\frac{\pi}{6}$
(red), with $\theta_{c}=\frac{\pi}{3}$ (green), and with $\theta_{c}=\frac{\pi}{2}$
(blue). Inset: Side view of two neighboring triangles $I$ and $J$
with their corresponding unit length normals $\vec{n}_{I}$ and $\vec{n}_{J}$.
\label{fig:critical_bending}}
\end{figure}

Some of the faceted shapes observed in experiments, as well as those
obtained in simulations (e.g. Fig.~\ref{fig:Snapshots-of-faceted}),
show sharp ridges with a large dihedral angle between neighboring
facets. Once the radius of curvature becomes comparable to the molecular
length scales the deformation is no longer locally small and microscopic
details of the vesicle start to play a role. Eq.~(\ref{eq:discrete_bending})
was derived\cite{Seung88} with the assumption that the curvature is
smooth and slowly varying between neighboring points, enabling a perturbative 
expansion of the bending energy in terms of small deviations of the normal
vectors to a planar surface. For a locally large deformation this is no longer 
justifiable and the expression for the bending energy needs to be modified. 
This is clearly a very hard problem that requires detailed knowledge of the microscopic
structure of the vesicle shell. Rather than constructing such a detailed
model we make the simple assumption that the bending energy saturates beyond 
a critical angle. In other words, Eq.~(\ref{eq:discrete_bending}) is modified to 
\begin{equation}
\tilde{E}_{bend}^{crit.}=\tilde{\kappa}\sum_{\left\langle I,J\right\rangle }\left(1-F\left(\theta_{IJ}\right)\right),\label{eq:discrtete-bending-critical-1}
\end{equation}
where 
\begin{equation}
F\left(\theta\right)=\left\{ \begin{array}{ccc}
\cos\left(\theta\right) & \mathrm{for} & \theta\leq\theta_{c}\\
\cos\left(\theta_{c}\right) & \mathrm{for} & \theta>\theta_{c}
\end{array}\right.\label{eq:discrete-benidng-critical-2}
\end{equation}
and, as in Eq.~(\ref{eq:discrete_bending}), $\theta_{IJ}$ measures
the angle between normals to a pair of neighboring triangles and $\theta_{c}$
is the critical value of that angle beyond which the bending penalty
saturates. A plot of $\tilde{E}_{bend}^{crit.}$ vs.~$\theta$ is
shown in Fig.~\ref{fig:critical_bending}. In this model $\theta_{c}$
is treated as an input parameter. In essence, $\theta_{c}$ controls
the deformation angle beyond which the vesicle shell is sufficiently
deformed that atomistic details start to be important. We assume that
the vesicle is homogenous sphere with uniform values for $\tilde{\kappa}$
and $\varepsilon$, and a standard triangular lattice with twelve
five-fold defects positioned at the corners of an inscribed icosahedron.
While one could argue that the stretching energy should also be modified by 
introducing a critical edge length beyond which the energy is constant,  we use, 
for simplicity, the expression in Eq.~(\ref{eq:discrete_stretch}) and note that test simulations show that the results are not qualitatively affected by this simplification.

As before, the low energy states were found using simulated annealing
Monte Carlo simulations. No constraints on volume or area were imposed.
Minimum energy configurations were found after $\approx10^{6}$ Monte
Carlo moves using a linear cooling protocol.

\begin{figure}
\begin{centering}
\includegraphics[scale=0.5]{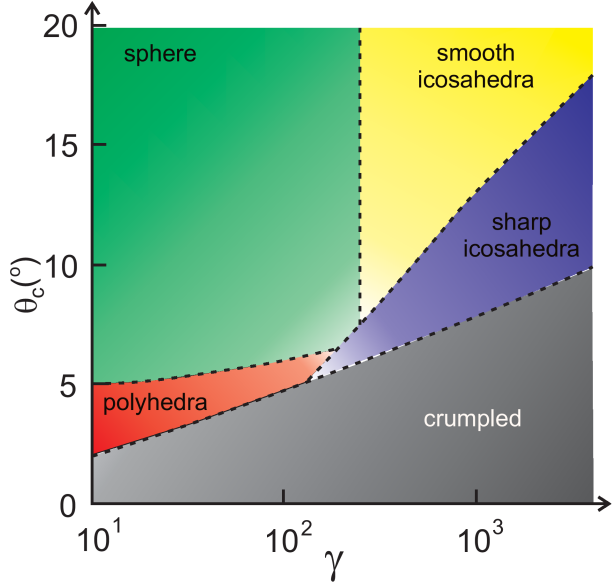} 
\par\end{centering}

\caption{Phase diagram for a vesicle with critical bending angle as a function
of the FvK number ($\gamma$) and the value of the critical angle
($\theta_{c}$, shown in degrees). We identify five distinct regimes.
In the ``sphere'' (green) and ``smooth icosahedra'' (yellow) regimes
$\theta_{c}$ is large and does not affect the shape. For small $\gamma$
the vesicle is spherical and crosses over into a smooth icosahedron
as $\gamma$ is increased. At lower values of $\theta_{c}$ the presence
of a critical angle starts to play a role and the vesicle shape changes
substantially. For small $\gamma$ the elastic energy is completely dominated
by the bending contribution and vesicles take irregular polyhedral shapes (red region). 
For larger values of $\gamma$ the stretching energy caused by topological 
defects cannot be ignored and the icosahedral symmetry reemerges, but 
this time accompanied by sharp ridges (blue region).
Finally, for very small values of $\gamma$ the vesicle is unstable to
crumpling (gray). \label{fig:Phase-diagram}}
\end{figure}

\begin{figure}
\begin{minipage}[t]{0.49\columnwidth}%
\begin{center}
\includegraphics[scale=0.1]{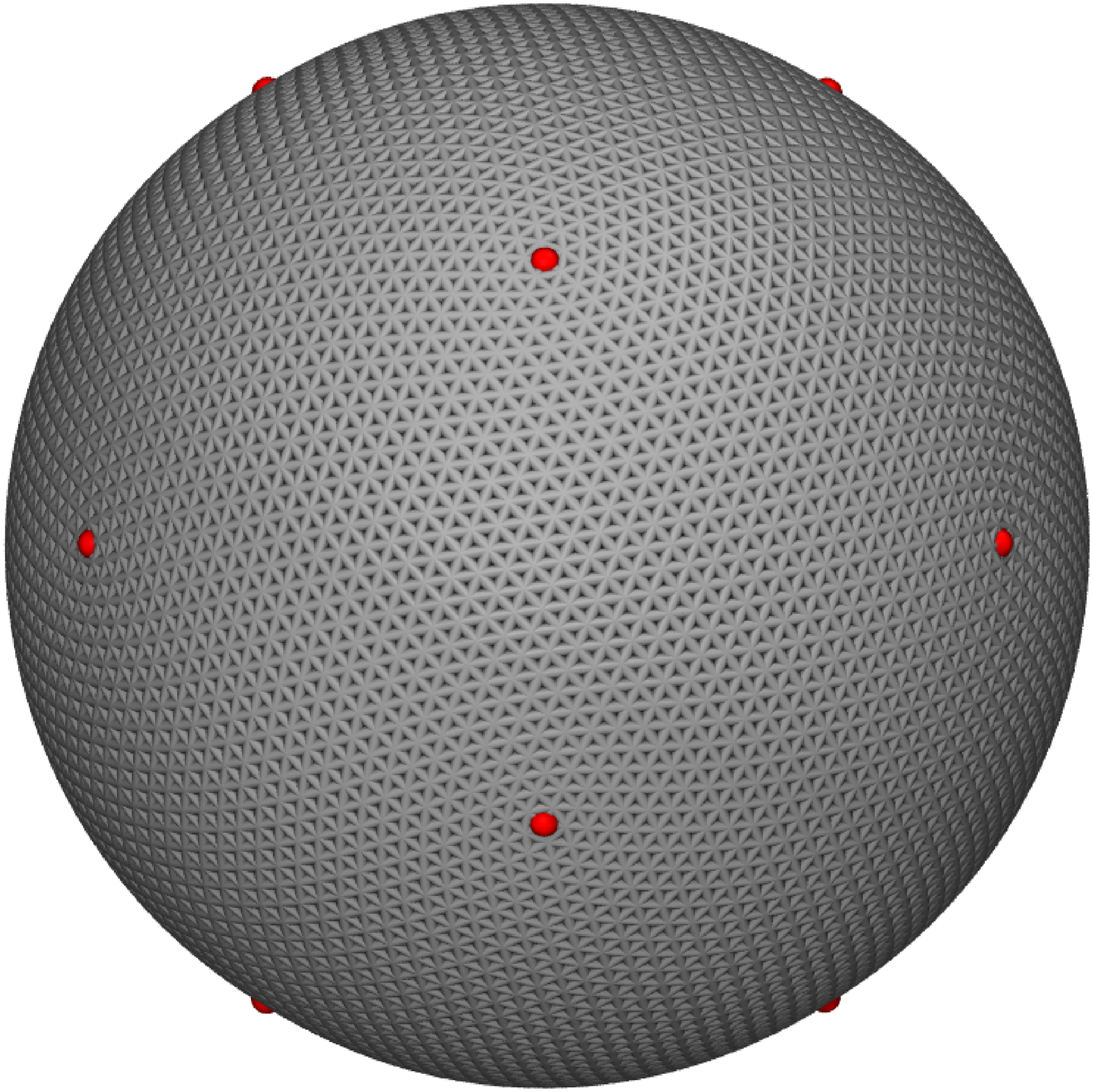} 
\par\end{center}

\begin{center}
(a) 
\par\end{center}%
\end{minipage}%
\begin{minipage}[t]{0.49\columnwidth}%
\begin{center}
\includegraphics[scale=0.1]{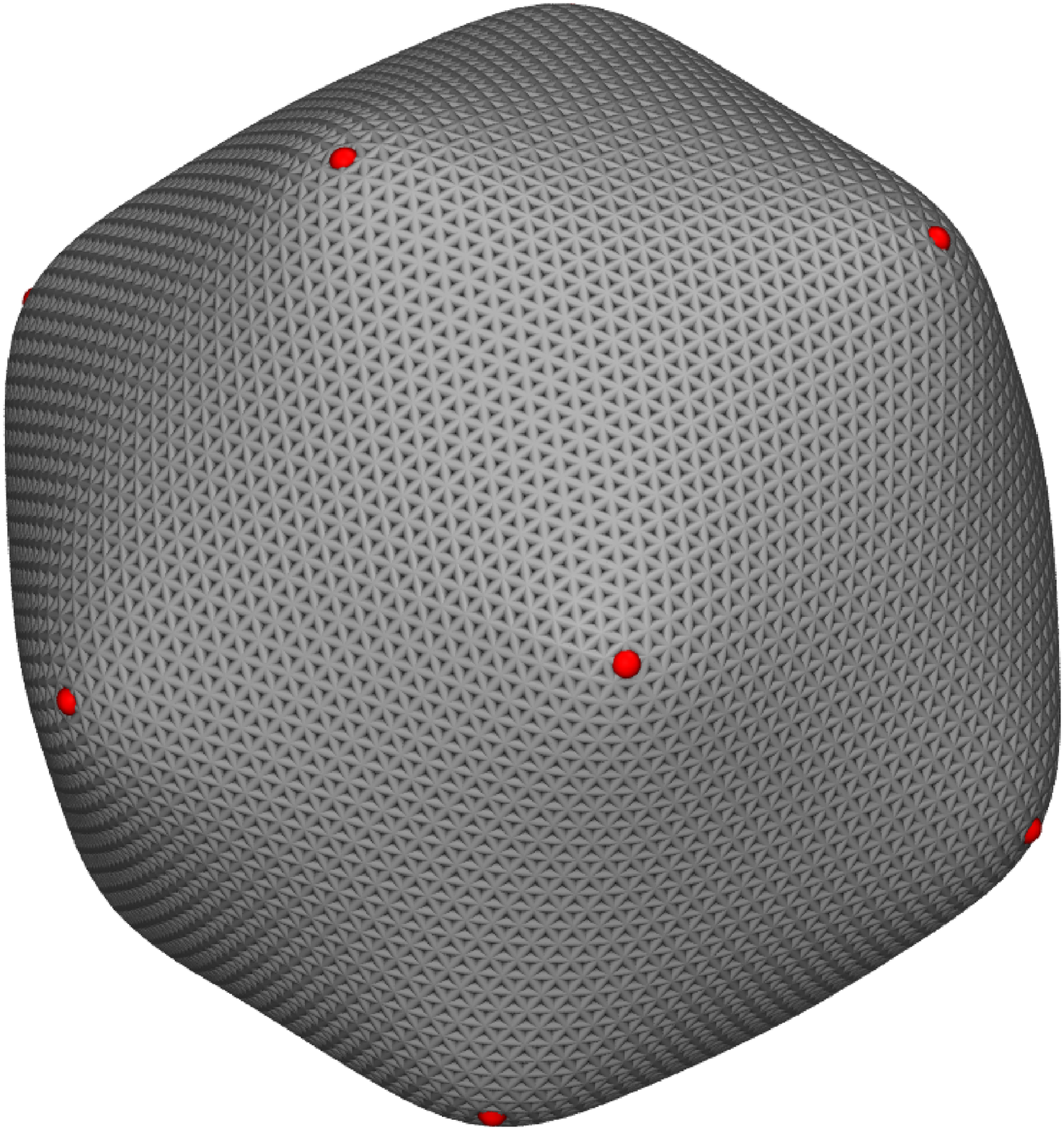} 
\par\end{center}

\begin{center}
(b) 
\par\end{center}%
\end{minipage}

\begin{minipage}[t]{0.49\columnwidth}%
\begin{center}
\includegraphics[scale=0.12]{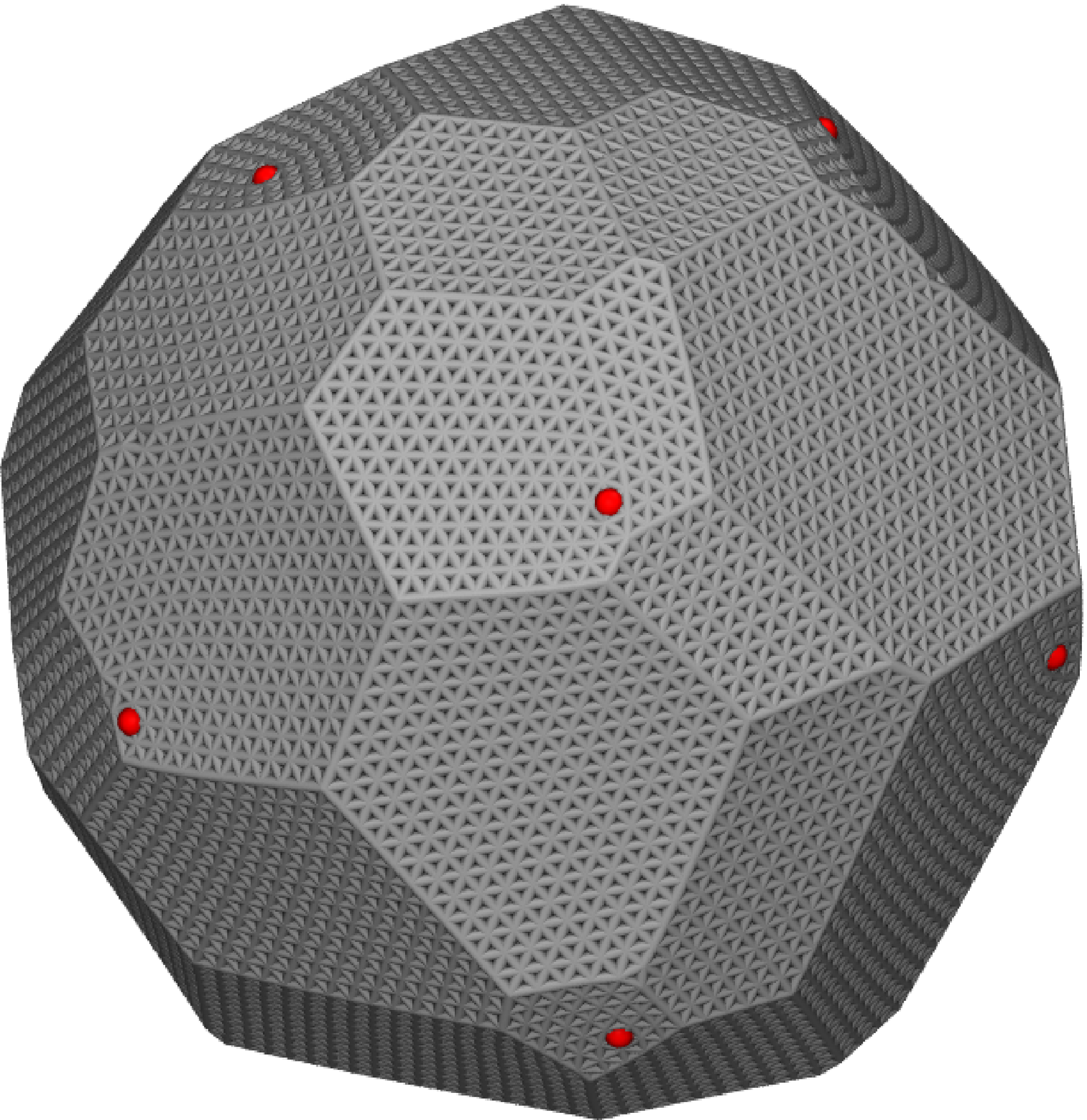} 
\par\end{center}

\begin{center}
(c) 
\par\end{center}%
\end{minipage}%
\begin{minipage}[t]{0.49\columnwidth}%
\begin{center}
\includegraphics[scale=0.15]{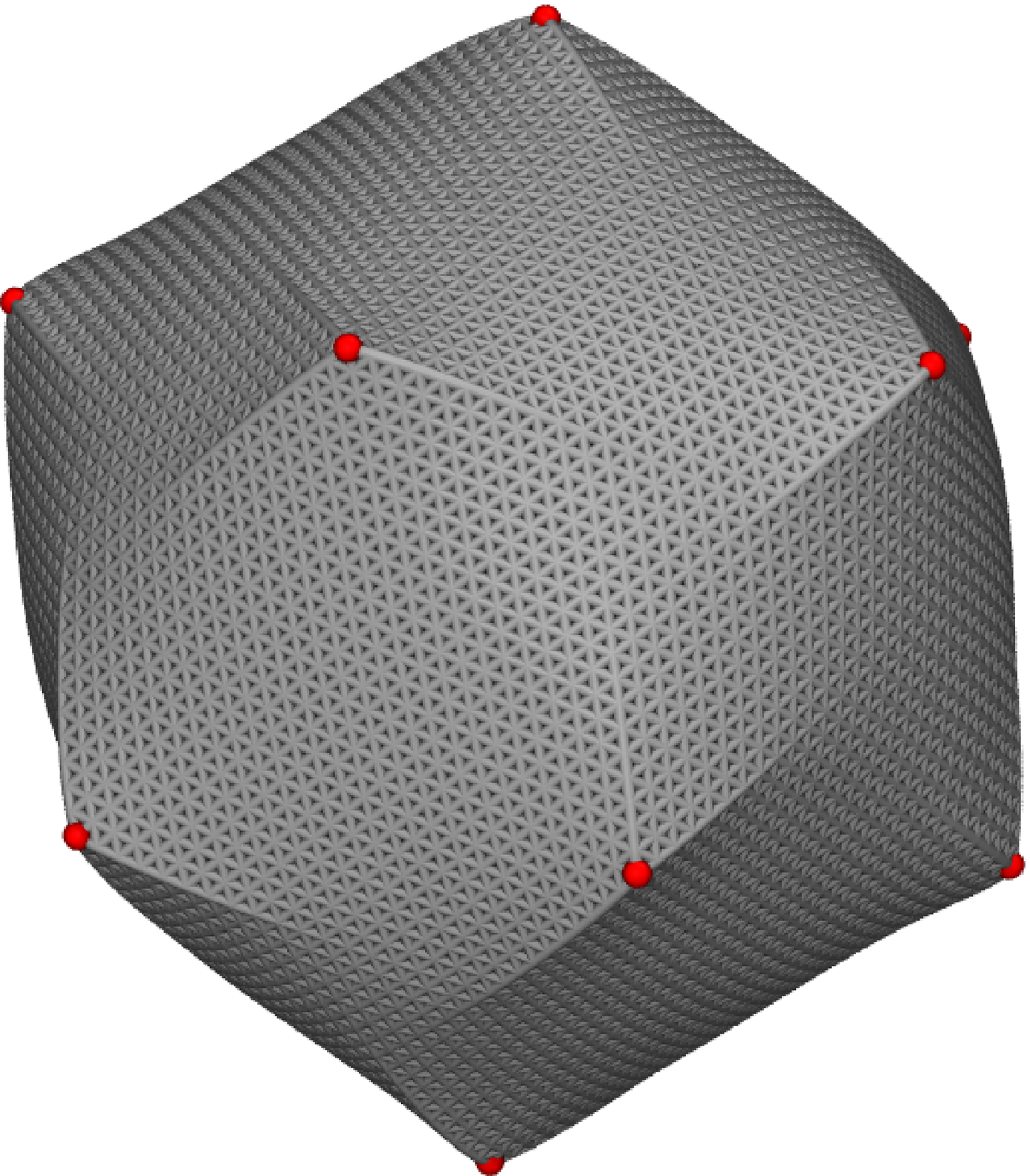} 
\par\end{center}

\begin{center}
(d) 
\par\end{center}%
\end{minipage}

\begin{centering}
\begin{minipage}[t]{0.49\columnwidth}%
\begin{center}
\includegraphics[scale=0.12]{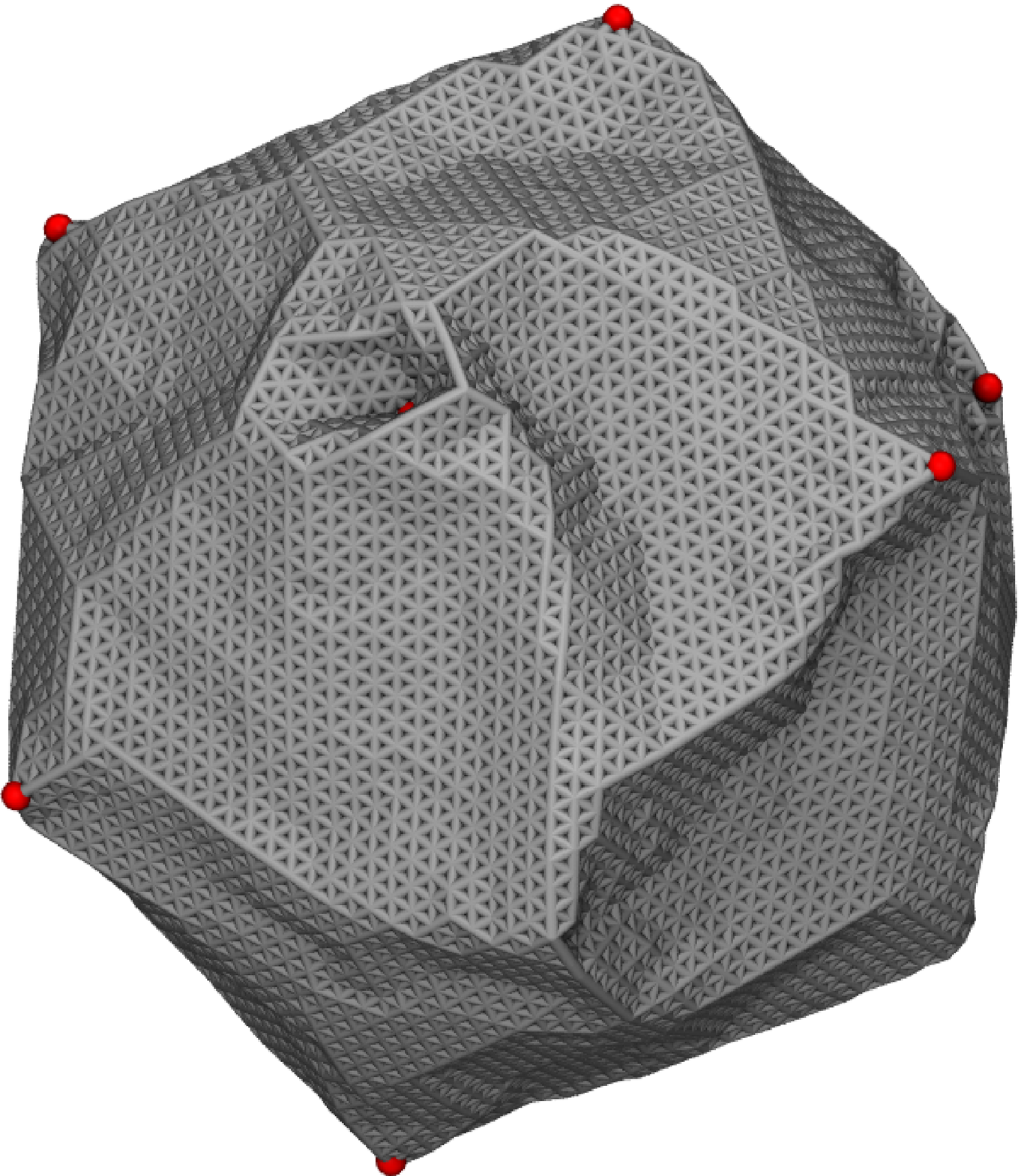} 
\par\end{center}

\begin{center}
(e) 
\par\end{center}%
\end{minipage}
\par\end{centering}

\caption{Snapshots of the low energy configurations of a vesicle constructed
using $\left(p,q\right)=\left(18,10\right)$ $T-$number triangulation
with $\approx6\times10^{3}$ vertices and radius $R\approx20.5l_{0}$.
(a) For low a FvK number $\gamma\approx50$ and a large critical angle
$\theta_{c}=20^{\circ}$the vesicle remains spherical. (b) For $\gamma\approx10^{3}$
and $\theta_{c}=20^{\circ}$ one observes a smooth buckled structure
with icosahedral symmetry akin to the shapes discussed in Section
\ref{sub:icosahedra}. (c) For small $\gamma\approx60$ and small
$\theta_{c}=5^{\circ}$ we find an irregular polyhedral vesicle with
facets being insensitive to the position of the defects. (d) As FvK
number if increased to $\gamma\approx10^{3}$ and the critical angle
set to $\theta_{c}=14^{\circ}$ the structure is icoshahedral but
with sharp ridges emanating for each defect. (e) Finally, at $\gamma\approx2.5\cdot10^{3}$
and very small critical angle, $\theta_{c}=2^{\circ}$ the vesicle
is no longer stable and crumples. Red points indicate positions of
twelve five-fold defects. \label{fig:snapshots-critical-angle}}
\end{figure}

We studied vesicles with approximately $6\cdot10^{3}$ vertices 
($\left(p,q\right)=\left(18,10\right)$-chiral; $\left(25,0\right)$-achiral) over a 
range of values of the FvK number ($\gamma$) and critical angle ($\theta_{c}$). 
In Fig.~\ref{fig:Phase-diagram} we display a phase diagram of this system mapped
with $\approx200$ independent pairs of $\gamma$ and $\theta_{c}$.
Five distinct regimes are identified. For low values of $\gamma$
and moderate to large $\theta_{c}$ (green region, denoted as ``sphere''
in Fig.~\ref{fig:Phase-diagram}) the bending penalty is too large and
vesicles remain spherical. In this region, the angles between
normals on any pair of neighboring triangles are smaller than $\theta_{c}$
and the existence of an angle cutoff plays no role in determining the 
shape: we recover the situation discussed in Section \ref{sub:icosahedra}.
If $\gamma$ is increased, while keeping $\theta_{c}$ large, one crosses
over into the regime where icosahedral shapes are favorable. Since
no deformation exceeds $\theta_{c}$ the transition occurs at the
same value of $\gamma$ as in the case without the critical bending
angle (yellow region, denoted as ``smooth icosahedra'' in Fig.~\ref{fig:Phase-diagram}).
If $\theta_{c}$ is reduced the existence of an angle cutoff is seen to affect the low energy shapes.
 In the region where $\gamma$ is small (red region denoted as ``polyhedra'' in Fig.~\ref{fig:Phase-diagram}) 
one observes irregular polyhedra with sizes and positions of facets
and edges that are insensitive to the presence of the five-fold defects.
This is to be expected since for small $\gamma$ the bending penalty completely
dominates over stretching and defect-driven buckling into a cone is
suppressed. In this region the faceting mechanism is similar to buckling
of the soft component as discussed in the previous section. As $\gamma$
is increased topological defects start to play a role and the
system transitions to the regime where defects drive buckling. With a
relatively small value of $\theta_{c}$, however, the system can further
lower its energy by forming sharp edges that emanate from the defects.
This region is denoted as ``sharp icosahedra'' (blue in Fig.~\ref{fig:Phase-diagram}).
We note that the direction of the sharp edges appears to depend on
the details of the triangulation and thus is not universal. Finally, if $\theta_{c}$ is too
small the vesicle becomes unstable and crumples (gray region denoted
as ``crumpled'' in Fig.~\ref{fig:Phase-diagram}). In this region
the simulations often fail to converge and instead get locked in high-energy
states that often occur in the early stages of a simulation. In Fig.~\ref{fig:snapshots-critical-angle} 
we show typical low energy configurations
in each of the five regimes.

We point out that this is a toy model with very little ingredients of
an actual system. Therefore, all conclusions should be
taken as qualitative with very little to no quantitative bearing for real
experimental systems such as lipid vesicles. However, we believe that some
general qualitative conclusions can still be deduced. For example, it is natural
 to ask which experimental systems could follow bending mechanism described by this model.
We note that for the continuum model to be applicable the deformation
has to be small at molecular length scales. In a lipid vesicle, on the other hand,
the angular difference between the long axes of
two neighboring molecules should be sufficient (a few degrees) for
the non-linear effects to be appreciable. It is reasonable to expect
such conditions to be met for vesicles in the sub $100nm$ size
range as are most of those for which faceting has been reported.\cite{blaurock1979small,Greenfield2009,leung2012molecular} 

\section{Summary and conclusions}

We have shown that faceted structures can be low energy configurations
of elastic vesicles as a result of several different mechanisms.
Incompatibility between crystalline order and spherical topology implies topological defects and residual stress even in the ground state. The stress can be relieved, in part, by buckling
into cones seeded at the defects. The buckled structure typically inherits 
the icosahedral symmetry associated with the defects. This mechanism can 
qualitatively account for the observation that smaller icosahedral viruses (e.g. \emph{Polyomaviruses}\cite{sweet1960vacuolating})
are smooth (spherical) while larger viruses (e.g. \emph{Mimivirus}\cite{la2003giant})
are buckled. For viruses the size of each capsomere is sufficiently
large ($\sim10\mathrm{nm}$) that the capsid is a nearly perfect lattice
with $\sim10^{2}-10^{3}$ units and no additional defects beyond the
minimum twelve five-fold sites required by the topology. In the case of
lipid bilayer vesicles lipid molecules are much smaller (occupying
an area $\lessapprox0.5nm^{2}$) and each vesicle is assembled from
hundreds of thousands of lipids. If such a vesicle develops crystalline order
as a result of cooling or strong electrostatic correlations
it is much harder to form a perfect lattice and one instead expects
grain boundaries to form. This effect is enhanced by the spherical
topology which hinders formation of a uniform crystal to begin with.
In the grain boundary region molecules are disordered and may even
be in the liquid state. As a result, the shell is expected to be softer.
We have shown that, if one allows for the presence of a softer component
in an otherwise crystalline vesicle, the low energy configuration
can be regular and irregular polyhedra that are not icosahedra. The
soft component forms ridges that are highly bent and separate hard
flat facets. The faceting mechanism is quite different from the
buckling into an icosahedron seeded at the five-fold sites. Furthermore,
it is reasonable to expect that the bending response of the vesicle
shell is not linear and significantly changes if the local deformation
becomes large. We introduced a simple model that takes into account
such effects at the most basic level. Even with a such simplified
model we were able to show that a non-linearity in the bending response
can drastically affect the low energy shapes and lead to interesting
faceted structures. 

Understanding the mechanisms that lead to the faceting of vesicles goes
beyond the realm of an academic exercise and could have far-reaching
impact on nano-technology. Flat faces, for example, could be easier
to functionalize or otherwise chemically treat. Biochemical reactions
are expected to have different rates near flat compared to curved
surfaces, which could be utilized to construct highly sensitive transport
agents. With the ability to accurately control the shape of vesicles
it could be possible to engineer vesicles that would, e.g. selectively
target only particular agents. We hope this work will broaden the
interest in this emergent subject of modern nano-science. 

This work was supported by NSF grant DMR-0808812. RS would like to
thank M.~Demers, C.~Funkhouser, C.~Leung, M.~Olvera de la Cruz, L.~Palmer, 
B.~Qiao, and G.~Vernizzi for collaborations on various aspects
of the study of faceting in multicomponent vesicles as well as to
R.~Everaers who suggested to analyze the effects of a critical angle bending.
RS and MJB also thank the Soft Matter Program at Syracuse University for financial
support. 

%\bibliographystyle{apsrev} %the RSC's .bst file
%\bibliography{elasticity,faceting} %your .bib file

\end{document}